# Magnetically-induced ferroelectricity in the $(ND_4)_2[FeCl_5(D_2O)]$ molecular compound.

José Alberto Rodríguez-Velamazán,[1,2] Óscar Fabelo,[2*] Ángel Millán,[1] Javier Campo,[1] Roger D. Johnson,[3] Laurent Chapon.[2*]

[1] *Instituto de Ciencia de Materiales de Aragón (ICMA), CSIC – Universidad de Zaragoza, 50009 Zaragoza, Spain.*

[2] *Institut Laue-Langevin, 38042 Grenoble Cedex 9, France.*

[3] *Univ. Oxford, Dept Phys, Clarendon Lab, Oxford OX1 3PU, England*

The number of magnetoelectric multiferroic materials reported to date is scarce, as magnetic structures that break inversion symmetry and induce an improper ferroelectric polarization typically arise through subtle competition between different magnetic interactions. The $(NH_4)_2[FeCl_5(H_2O)]$ compound is a rare case where such improper ferroelectricity has been observed in a molecular material. We have used single crystal and powder neutron diffraction to obtain detailed solutions for the crystal and magnetic structures of $(NH_4)_2[FeCl_5(H_2O)]$, from which we determined the mechanism of multiferroicity. From the crystal structure analysis, we observed an order-disorder phase transition related to the ordering of the ammonium counterion. We have determined the magnetic structure below $T_N$, at 2K and zero magnetic field, which corresponds to a cycloidal spin arrangement with magnetic moments contained in the *ac*-plane, propagating parallel to the *c*-axis. The observed ferroelectricity can be explained, from the obtained magnetic structure, via the inverse Dzyaloshinskii-Moriya mechanism.

*This work is in honor of Prof. Fernando Palacio on the occasion of his retirement*

## Introduction

Multiferroic materials, compounds presenting at least two ferroic- orders simultaneously, have attracted considerable interest due to the possible cross control of physical properties. In particular, magnetoelectric multiferroics where an electrical polarization is induced by a magnetic transition, the so-called type-II multiferroics, have been extensively studied for their fundamental properties as well as for their potential applications in memories, sensors, transducers, etc.[1]

The complex magnetic ordered state in these improper (or pseudo-proper) ferroelectric materials [2] is stabilized by competitions between different exchange couplings which break inversion symmetry and induce a ferroelectric polarization. The most studied multiferroics of this type are the "cycloidal" magnets such as $RMnO_3$ [3] $MnWO_4$[4] or $CoCr_2O_4$ [5], where ferroelectricity emerges as a consequence of a cycloidal modulation of the magnetic moments. The coupling mechanism between magnetization and ferroelectricity has been explained by the so-called *inverse Dzyaloshinskii-Moriya* (DM) effect [6] or spin current mechanism, that predict that the electric polarization, **P**, is proportional to $\mathbf{r}_{ij} \times (\mathbf{S}_i \times \mathbf{S}_j)$, where $\mathbf{r}_{ij}$ is the vector connecting the nearest spins, $\mathbf{S}_i$ and $\mathbf{S}_j$, and $(\mathbf{S}_i \times \mathbf{S}_j)$ is the so-called spin-chirality vector.[7]

One of the approaches to the design of materials combining different properties is the association of different building blocks that carry the different properties of interest. When a ferroelectric material is wanted, the use of ammonium molecules substituting monovalent counterions (i.e. Na, K, Rb or Cs) in inorganic networks has been proved as a successful approach in order to induce a switchable ferroelectric transition. This is the case of peroxychromates of general formula $M_3CrO_8$,[8] where the substitution of the alkali metals by an ammonium molecule produces a system with a well-defined ferroelectric phase transition.[9] A similar approach has been used in metal-organic compounds, where the combination of two different networks with different properties has been shown to be an excellent strategy to design new multiferroic materials.[10] Nevertheless, in all these examples the existence of ferroelectric or antiferroelectric transitions is due to the occurrence of an order–disorder phase transition where the hydrogen bonds between the guest molecule and the host-framework play a central role. No interplay between the magnetic and electric orders has been observed in these cases, and therefore improper ferroelectricity has never been described for this type of compounds. However, a recently published work focused on ammonium pentacloroaquaferrate(III), with formula $(NH_4)_2[FeCl_5(H_2O)]$, has characterized this molecular compound as a new multiferroic material, where magnetic and electric properties are connected.[11] The heat capacity of the deuterated sample, of formula $(ND_4)_2[FeCl_5(D_2O)]$ (see Figure S1), shows a behavior equivalent to that of the hydrogenated sample reported by Ackermann *et al*,[11] presenting two closely spaced peaks in the low temperature region at ca. 6.9 and 7.2 K.[11,12] At high temperature (*ca* 79 K), this material reveals a third signal in the heat capacity, associated to a suspected structural phase transition, although the previous X-ray diffraction measurements above and below this transition did not show any significant changes.[11] The single crystal susceptibility measurements indicate an *XY* anisotropy with the *ac*-plane being the magnetic easy plane. The spontaneous electric polarization obtained integrating the pyroelectric current shows that the system becomes ferroelectric below $T_{FE}$ ~ 6.9K at zero magnetic field, corresponding with the low temperature cusp in the heat capacity. The polarization has two components, the main one along the *a*-axis with a value of $3\mu Cm^{-2}$ and a secondary one, with a value ten times lower, along the *b*-axis. Thus, the spontaneous electric polarization lies in the *ab*-plane.

In this paper, we investigate the temperature dependence of the crystal structure using X-ray and neutron diffraction measurements (on hydrogenated and deuterated samples respectively). Below $T_N$, (at 2 K) we have determined unambiguously the magnetic structure at zero magnetic field. A simple model based on a set of four exchange coupling interactions has been proposed to explain the magnetic ground state. We present also the temperature evolution of some selected magnetic reflections in order to evaluate the possible spin-reorientation above 6.9 K. Finally, we propose the possible multiferroicity mechanism in $(ND_4)_2[FeCl_5(D_2O)]$ compound.

## Experimental methods

### Materials

Deuterated reagents and solvents used in the synthesis were purchased from commercial sources and used without further purification. The synthesis of non-deuterated samples has been carried out with the same protocol described below but with hydrogenated reagents.

### Synthesis of $(ND_4)_2[FeCl_5(D_2O)]$

Single crystals of $(ND_4)_2[FeCl_5(D_2O)]$ of suitable size for neutron scattering were prepared by the seeded growth technique, by slow evaporation of saturated solutions at 50 ºC. The seeds were formed by cooling of hot saturated solutions. The resulting compound was directly checked by single crystal X-ray diffraction and the results were in agreement with the unit cell parameters and the crystal structure previously reported. [11-12]

### X-ray Single Crystal Refinements

Data collections of $(NH_4)_2[FeCl_5 \cdot H_2O]$ were carried out at room temperature (HT) and 50 K (LT) in an Agilent SuperNova X-ray µ-focus source equipped with a helium cryostream, using Mo-Kα radiation (λ = 0.71073 Å). The reflections were indexed, integrated and scaled using CrysAlis Pro program.[13] The structures of **1_HT** and **1_LT** were solved by direct methods using the SHELXS97 program at the space group *Pnma* (No. 62). All non-hydrogen atoms were refined anisotropically by full-matrix least-squares technique based on $F^2$ using SHELXL97.[14] The hydrogen atoms were positioned geometrically and refined using the difference electron density map applying DFIX soft-constrains, giving rise to very close models for both temperatures; therefore no clear signal of the structural phase transition was observed. The only observed change comparing the HT with the LT measurements is the cell volume compression, which is around 2 %, value which is compatible with the thermal expansion effect. The final geometrical calculations and the graphical manipulations were carried out with PARST97,[15] PLATON[16] and DIAMOND[17] programs.

### Neutron Diffraction Measurements.

Single-crystal neutron diffraction data were collected at the hot-neutrons four-circle diffractometer D9 at Institut Laue Langevin (ILL, Grenoble, France) with a wavelength of 0.8322(1) Å obtained with a copper (113) monochromator operating in transmission geometry. D9 diffractometer is equipped with a 2D detector of 6 x 6 cm (32 x 32 pixels) that allow us a reciprocal space survey. The crystal was mounted onto specific aluminum sample holders that produce a low background and sealed into a closed-cycle cryostat.

In order to determine the crystal structure above and below the nuclear phase transition (79 K) data collections consisting in a combination of omega- and omega-2theta-scans of each individual reflection where carried out at 100 and 2 K, yielding 4900 and 2710 independent reflections respectively. The cell parameters where obtained at both temperatures and the crystal structure refined at 100 K and accurately determined below the phase transition. Due to the occurrence of a magnetic propagation vector different of **k** = (0,0,0), nuclear and magnetic reflections are not overlapped below the magnetic phase transition ($T_N$ ca 7.25 K) and therefore both the magnetic and nuclear phases can be measured independently. The magnetic structure was determined at 2 K from 127 independent reflections. Additionally, the evolution of two different magnetic reflections [(0,0,-1-$k_z$) and (0,-3, $k_z$)] was followed in the temperature range from 2 to 12 K. The main crystallographic data are summarized in Table 1.

**Table 1.** Experimental parameters and main structural crystallographic data for the studied compounds.

| Formula | [ND$_4$]$_2$[Fe$^{III}$Cl$_5$(D$_2$O)] | |
|---|---|---|
| Empirical Formula | Cl$_5$D$_{10}$FeN$_2$ | |
| M$_r$ (g·mol$^{-1}$) | 297.26 | |
| Temperature (K) | 100(2) | 2(2) |
| Inst. λ(Å) | D9, 0.83220 | D9, 0.83220 |
| Crystal system | Orthorhombic | Monoclinic |
| Space group (No.) | *Pnma* (62) | *P*112$_1$/*a* (14) |
| Crystal size (mm) | 3 × 3 × 3 | 3 × 3 × 3 |
| *a* (Å) | 13.5221(9) | 13.5019(7) |
| *b* (Å) | 9.9305(6) | 9.9578(5) |
| *c* (Å) | 6.9219(6) | 6.9049(4) |
| α(°) | 90.00 | 90.00 |
| β(°) | 90.00 | 90.00 |
| γ(°) | 90.00 | 90.109(4) |
| V (Å$^3$) | 929.48(12) | 928.36(9) |
| Z | 4 | 4 |
| ρ$_c$ (g·cm$^{-3}$) | 2.124 | 2.127 |
| Meas. Reflections/ (R$_{int}$) | 2710 (0.0270) | 4900 (0.0234) |
| Indep. ref. [I > 2σ(I)] | 1328 | 1673 |
| Parameters/restraints. | 132/ 0 | 174 / 0 |
| Hydrogen treatment | Refall | Refall |
| Goodness of fit | 1.111 | 0.976 |
| Final R indices [I > 2σ(I)]: R$_1$ / wR$_2$ | 0.0431/0.0980 | 0.0402/0.0966 |
| R indices (all data): R$_1$ / wR$_2$ | 0.0484/0.1010 | 0.0605/0.0988 |

*The crystallographic details of the data collections at RT and 50 K using single crystal X-ray diffraction have been provided in the Supplementary Information (see text and Table S1).

The program RACER[18] was used to integrate the omega- and omega-2theta-scans and to correct them for the Lorentz factor. The crystal attenuation corrections were performed with a prism model using DATAP program,[19] with an estimated total neutron absorption coefficient of 0.057 cm$^{-1}$.[20] Additional high-resolution neutron powder diffraction measurements at D2B instrument (ILL) were performed at 45 K in order to verify the structural transformation at 79 K. The sample was contained in a Ø 6 mm cylindrical vanadium-can inside an Orange Cryostat. The neutron diffraction pattern was collected using 1.5642 Å wavelength.

The refinements of the nuclear structures were performed using the programs SHELX[14] and FullProf Suite.[21] The magnetic structure was treated with the program FullProf using the integrated intensities obtained from the single-crystal measurements. The models for the magnetic structure were deduced from the symmetry analysis techniques implemented in the program BasIreps included in the FullProf Suite.[21] The magnetic moments were localized on the Fe(III) atoms and, in the final refinement, the real and imaginary parts of the Fourier coefficients of the magnetic moments were constrained to have the same magnitude with directions described by spherical angles. The nuclear and magnetic contributions to the diffraction intensities were treated as two separate patterns, with the magnetic phase described in space group *P*-1 (No. 2), using the magnetic form-factor curve of Fe(III) and the scale factor obtained from the refinement of the nuclear phase at the same temperature.

## Results and discussion
### Neutron Studies

**Crystal structure at 100 K**

The title compound is built up from one [FeCl$_5$(D$_2$O)]$^{2-}$ ion and two ND$_4$ counterions, and the crystal structure is held together by an extensive network of H-bonds in concurrence with the ionic interaction (see Figure 1). The KPI packing index gives a percent of filled space of 81.5,[22] therefore there is no accessible space in the network for solvents. The crystallographic properties of the two phases (above and below the structural phase transition), including space groups, lattice parameters as well as some details of the data refinement, are shown in Table 1.

**Table 2.** Main structural variations between high and low temperature phases obtained from single crystal neutron diffraction refinement.

| 100K | | 2K | |
|---|---|---|---|
| Distances (Å) | | Distances (Å) | |
| Fe(1)-O(1W) | 2.095(2) | Fe(1)-O(1w) | 2.095(2) |
| Fe(1)-Cl(1) | 2.3155(15) | Fe(1)-Cl(1) | 2.391(2) |
| Fe(1)-Cl(2) | 2.3926(7) | Fe(1)-Cl(2) | 2.400(2) |
| Fe(1)-Cl(2)$^a$ | 2.3926(7) | Fe(1)-Cl(3) | 2.3375(14) |
| Fe(1)-Cl(3) | 2.3435(16) | Fe(1)-Cl(4) | 2.3179(13) |
| Fe(1)-Cl(4) | 2.3892(16) | Fe(1)-Cl(5) | 2.4026(14) |
| N(1)-D(1A) | 1.035(6) | N(1)-D(1) | 1.027(3) |
| N(1)-D(2A) | 0.959(5) | N(1)-D(2) | 1.007(4) |
| N(1)-D(3A) | 1.013(5) | N(1)-D(3) | 1.004(2) |
| N(1)-D(4A) | 1.015(5) | N(1)-D(4) | 1.026(3) |
| N(1)-D(1B) | 1.026(3) | N(2)-D(5) | 1.012(3) |
| N(1)-D(2B) | 1.019(5) | N(2)-D(6) | 1.027(4) |
| N(1)-D(3B) | 1.008(4) | N(2)-D(7) | 1.026(3) |
| N(1)-D(4B) | 0.970(4) | N(2)-D(8) | 1.018(3) |
| O(1w)-D(1w) | 0.958(2) | O(1W)-D(1W) | 0.963(4) |
| | | O(1W)-D(2W) | 0.973(4) |
| Angles (°) | | Angles (°) | |
| Cl(1)-Fe(1)-Cl(2) | 94.40(1) | Cl(1)-Fe(1)-Cl(2) | 170.73(1) |
| Cl(1)-Fe(1)-Cl(2)$^a$ | 94.40(1) | Cl(1)-Fe(1)-Cl(3) | 90.53(1) |
| Cl(1)-Fe(1)-Cl(3) | 90.49(1) | Cl(1)-Fe(1)-Cl(4) | 94.78(1) |
| Cl(1)-Fe(1)-Cl(4) | 89.86(1) | Cl(1)-Fe(1)-Cl(5) | 89.61(1) |
| Cl(1)-Fe(1)-O(1W) | 179.76(1) | Cl(1)-Fe(1)-O(1W) | 85.51(1) |
| Cl(2)-Fe(1)-Cl(2)$^a$ | 171.08(1) | Cl(2)-Fe(1)-Cl(3) | 91.23(1) |
| Cl(2)-Fe(1)-Cl(3) | 90.70(1) | Cl(2)-Fe(1)-Cl(4) | 94.29(1) |
| Cl(2)-Fe(1)-Cl(4) | 89.28(1) | Cl(2)-Fe(1)-Cl(5) | 88.55(1) |
| Cl(2)-Fe(1)-O(1W) | 85.60(1) | Cl(2)-Fe(1)-O(1W) | 85.40(1) |
| Cl(3)-Fe(1)-Cl(2)$^a$ | 90.70(1) | Cl(3)-Fe(1)-Cl(4) | 90.74(1) |
| Cl(3)-Fe(1)-Cl(4) | 179.64(1) | Cl(3)-Fe(1)-Cl(5) | 179.44(1) |
| Cl(3)-Fe(1)-O(1W) | 89.75(1) | Cl(3)-Fe(1)-O(1W) | 89.87(1) |
| Cl(4)-Fe(1)-Cl(2)$^a$ | 89.28(1) | Cl(4)-Fe(1)-Cl(5) | 89.78(1) |
| Cl(4)-Fe(1)-O(1W) | 89.90(1) | Cl(4)-Fe(1)-O(1W) | 179.32(1) |
| O(1W)-Fe(1)-Cl(2)$^a$ | 85.60(1) | Cl(5)-Fe(1)-O(1W) | 89.60(1) |
| D(1W)-O(1W)-D(1W)$^a$ | 110(1) | D(1W)-O(1W)-D(2W) | 109(1) |

Symmetry codes: (a) *x*, 0.5-*y*, *z*

The Fe(III) atom is located on a mirror plane perpendicular to the *b*-axis, and presents a slightly distorted octahedral environment. The shortest Fe-Cl bond distances correspond to Cl(1), which is coordinated in *trans*- conformation with respect to the water molecule (see Table 2). The interbond angles in the anion are slightly different of those of an ideal octahedron. These deviations are related with the strength of the different hydrogen bonds present in this compound. A clear example is the coordination water molecule hydrogen bond [O(1w)-D···Cl(2)], with a D···Cl distance of 2.217(2) Å. As result of this interaction the Fe-Cl(2) bond distances are slightly elongated. The influence of the H-bonds due to the $ND_4$ counterion is more difficult to quantify due to the extension of the H-bond network in which the $ND_4$ molecule is involved. A detailed list with all the possible H-bonds can be consulted in Table 3.

**Table 3.** Relevant hydrogen bonds for 100 K (**1_HT**) and 2 K (**1_LT**) measurements obtained from single crystal neutron diffraction refinement.

| D-H···A[*] | H···A / Å | D-H···A / ° |
|---|---|---|
| 100K | | |
| O(1W)-D(1W)···Cl(2)[b] | 2.217(2) | 178.39(14) |
| N(1)-D(1A)···Cl(2)[c] | 2.540(6) | 173.5(4) |
| N(1)-D(2A)···Cl(4)[d] | 2.364(6) | 160.8(4) |
| N(1)-D(3A)···Cl(4) | 2.368(5) | 171.3(4) |
| N(1)-D(4A)···Cl(2)[e] | 2.484(6) | 160.1(4) |
| N(1)-D(1B)···Cl(2)[d] | 2.466(3) | 173.6(2) |
| N(1)-D(2B)···Cl(2)[a] | 2.418(5) | 164.7(4) |
| N(1)-D(3B)···Cl(1)[f] | 2.329(4) | 169.7(3) |
| N(1)-D(4B)···Cl(3)[g] | 2.355(4) | 159.6(3) |
| 2K | | |
| O(1w)-D(1w)···Cl(2)[h] | 2.217(4) | 175.8(3) |
| O(1w)-D(2w)···Cl(1)[i] | 2.211(4) | 178.3(3) |
| N(1)-D(1)···Cl(5) | 2.314(3) | 173.3(2) |
| N(1)-D(2)···Cl(5)[j] | 2.334(4) | 159.4(3) |
| N(1)-D(3)···Cl(1)[g] | 2.454(2) | 178.3(2) |
| N(1)-D(4)···Cl(1)[f] | 2.399(3) | 161.4(2) |
| N(2)-D(5)···Cl(2)[h] | 2.391(2) | 175.8(2) |
| N(2)-D(6)···Cl(3) | 2.286(4) | 162.0(3) |
| N(2)-D(7)···Cl(2)[k] | 2.374(3) | 163.8(2) |
| N(2)-D(8)···Cl(4)[l] | 2.312(3) | 169.2(3) |

[*] D and A stand for donor and acceptor, respectively.
Symmetry codes: (a) *x*, 0.5-*y*, *z*; (b) 2-*x*, -0.5+*y*, -*z*; (c) *x*, 0.5-*y*, 1+*z*; (d) 2-*x*, -0.5+*y*, 1-*z*; (e) 1.5-*x*, -0.5+*y*, 0.5+*z*; (f) 1.5-*x*, -*y*, 0.5+*z*; (g) *x*, *y*, 1+*z*; (h) 2-*x*, 1-*y*, -*z*; (i) 2-*x*, -*y*, -*z*; (j) 2-*x*, -*y*, 1-*z*; (k) *x*, *y*, -1+*z*; (l) 1.5-*x*, 1-*y*, -0.5+*z*

The refinement of the $ND_4$ molecule at high temperature was made by several successive attempts. Our first model considered the $ND_4$ molecule ordered. An initial model was obtained using soft constraints, with the $ND_4$ molecule geometrically fixed and only small deviations of this ideal geometry allowed. A similar protocol was also applied using rigid bodies in FullProf suite [21] but in that case the $ND_4$ molecule forms an undeformable rigid body with a perfect tetrahedral symmetry. In order to define a realistic model, the N-D bond length was fixed at 1.025 Å and the D-N-D angle was fixed to 109°, values in agreement with other ammonium-containing compounds. [23] The values of the Fe–Cl, Fe-O and O-D bond lengths were left variable. Unfortunately all our attempts to refine this model produce unstable solutions. A more complex model considering the $ND_4$ molecule disordered was subsequently tested. For the starting point of this model, the D atoms of the $ND_4$ molecule were set in 8 different sites (Wyckoff position 8*d*) with an occupancy of 0.5 that is equivalent to two disordered $ND_4$ units with half occupancy.

The first set of refinements was made using the Simulated Annealing procedure included on the FullProf suite.[21] The refinement produced a solution where the 8 D atoms around the N can be seen as two superimposed tetrahedrons (see Figure 1). This model was the starting point for the latter refinement using the SHELX program.[14] The final refinement was made without restraints. The values of the N-D bond distances as well as the D-N-D angles are in agreement with those observed in other ammonium-containing compounds.[23]

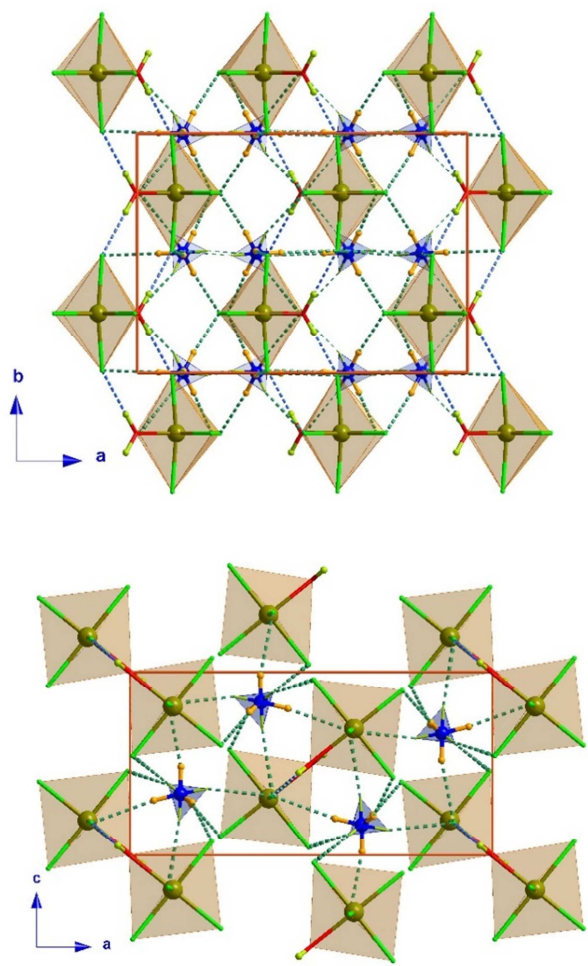

**Figure 1.** Views of the 100K crystal packing along the *c*-axis (top) and *b*-axis (bottom) of $(ND_4)_2[FeCl_5(D_2O)]$ compound, where the O-D···Cl and the N-D···Cl hydrogen bonds have been represented as blue and green dashed lines, respectively. For the sake of clarity only one conformation of the disordered $ND_4$ molecule has been drawn in polyhedron representation.

The deuterium atoms of the disordered ND$_4$ act as donors in an extensive hydrogen-bond network where each deuterium is involved in a single H-bond, with distances ranging from 2.329(4) to 2.540(6) Å. The shortest H-bonds are those involving the coordinated water molecule and both *trans*-Cl(2) atoms, which connect the Iron(III) atoms forming a zig-zag chain along the *b*-axis.

The shortest Fe⋯Fe distance connects iron atoms in the Fe(1)-O(1w)-D⋯Cl(2)-Fe(1) zig-zag chain running along the *b*-axis, with a distance between metallic centers of 6.457(1) Å, while the shortest interchain distance connects the Fe atoms within the *ac*-plane, with a distance of 6.8125(14) Å.

**Crystal structure determination at low temperature**

There have been several previous attempts to refine the crystal structure of this compound at low temperature (below the structural phase transition);[11,12,24] nevertheless the reported structures above and below the phase transition were mainly the same, and neither changes in the space group nor a significant displacement on the atomic positions were observed. These observations have been also verified in this work with X-ray single crystal diffraction data collections above and below the nuclear phase transition. The best model fitting the X-ray data is always obtained with the structure refined on the space group *Pnma*, independently of the temperature (in our case RT and 45 K, see Supplementary Information for details). These results suggest that the hydrogen atoms present in the structure are responsible for the phase transition, and therefore X-ray diffraction would not be the best probe to analyze it due to the low scattering power of X-rays by hydrogen. In order to verify our hypothesis, neutron diffraction measurements of the deuterated compound have been carried out on a powder sample (at 45 K) and on a single crystal (at 100 and 2K).

Assuming that the possible phase transition is a second order ("continuous") phase transition, we determined the list of possible subgroups of the high temperature space group (*Pnma*). The only possible space groups are ***Pna2$_1$*** (No. 33), ***Pmn2$_1$*** (No. 31), ***Pmc2$_1$*** (No. 26), $P2_12_12_1$ (No. 19), $P2_1/c$ (No. 14), $P2_1/m$ (No. 11), ***Pc*** (No. 7), ***Pm*** (No. 6), ***P2$_1$*** (No. 4), *P*-1 (No. 2) and *P* (No. 1), where the polar ones are written in bold. Given that the structural transition (ca 79 K) is not associated with the onset of a ferroelectric state,[11] the only possible space groups immediately below this phase transition are in principle the non-polar ones. Once the system becomes ferroelectric below 6.9 K, these non-polar groups are no more strictly correct, but the structural changes responsible of this type of electric polarization are usually too subtle to be observable,[4] so we assumed the same space group at 45 K and 2 K. All possible space groups were tested by trial and error and the only subgroup that refines properly the experimental data was found to be $P2_1/c$ (No. 14). However, in order to facilitate comparisons between the high- (100K) and low-temperature phases the refinement was carried out in the $P112_1/a$ space group, which is a non-standard setting of $P2_1/c$ (No. 14). The reflection splitting due to the subtle monoclinic distortion can be clearly observed in the high resolution neutron powder diffraction data at 45K (see Figure S2).

The results of the refinement of the single crystal data show a crystal structure at low temperature which is built up from one and two crystallographically independent [FeCl5·H2O]$^{2-}$ and ND$_4$ units, respectively (see Figure 2). As previously mentioned, the low temperature phase crystallizes in $P112_1/a$ space group, which is a maximal non-isomorphic subgroup of *Pnma* with index 2. Therefore the number of crystallographically independent atoms at low temperature is doubled with respect to the high temperature phase. The phase transition involves the loss of a *n*-glide parallel to the *a*-axis and a mirror plane which is contained in the *ac*-plane, together with a slight variation in the cell parameters and the cell volume (see Table 1). The KPI packing index gives a percent of filled space of 77.2, a value which is slightly lower than that observed at high temperature.[22] However, as occurs in the high-temperature phase, there is no accessible space in the network for solvents.

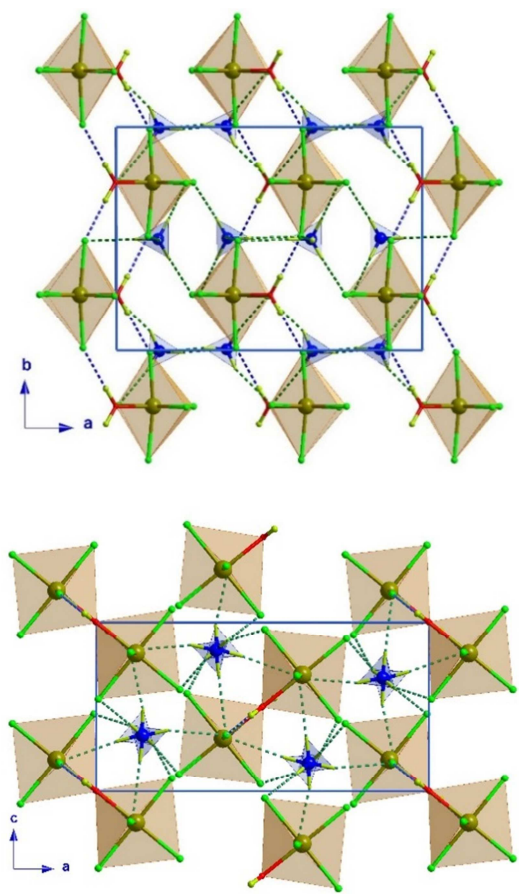

**Figure 2.** Views of the 2K crystal packing along the *c*-axis (left) and *b*-axis (right) of (ND$_4$)$_2$[FeCl$_5$(D$_2$O)] compound, where the O-D⋯Cl and the N-D⋯Cl hydrogen bonds have been represented as blue and green dashed lines, respectively

At low temperature, the Fe(III) atoms are located on a *4e* general position due to the loss of the mirror plane perpendicular to the *b*-axis. The Fe(III) environment is slightly more distorted compared with the high temperature structure, but the octahedral environment is held. The largest variations are observed in the long bond distances involving Cl(1), Cl(2) and Cl(5), where a significant elongation of the Fe-Cl distance has been identified (see Table 2). The interbond angles in the anion are very similar to those observed at high temperature (see Table 2).

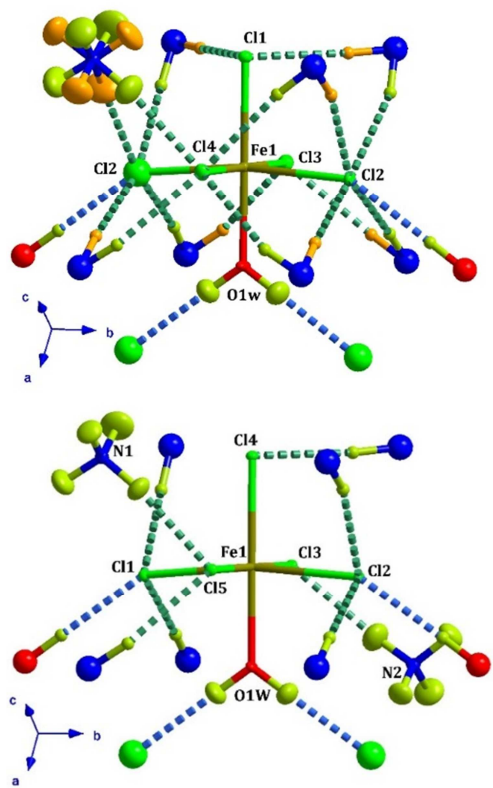

**Figure 3.** View of the hydrogen bond interactions between the [FeCl$_5$(D$_2$O)] unit and the adjacent molecules obtained from the neutron diffraction crystal structure at 100 (top) and 2 K (bottom). The O-D···Cl and the N-D···Cl hydrogen bonds have been represented as blue and green dashed lines, respectively. A detailed list of relevant hydrogen bond distances and angles can be consulted in Table 3. The asymmetric units are given in ORTEP representation with atoms at 50 % of probability. Ammonium hydrogen atoms in the HT phase, have been drawn on lime and orange colors in order to denote the two different conformations.

The most remarkable difference between high- and low-temperature phases concerns the ND$_4$ counterions which are well ordered below the phase transition (ca 79 K). An order-disorder transition is therefore proposed as the origin of this structural phase transition. The influence of the H-bond network in this transition has been taken into account by analyzing the bond-distances and the bond-orientation. At low-temperature all the bond-distances are slightly shorter than those observed at high-temperature. The bond-orientations are very similar, and in both cases the donor atoms interact with only one acceptor, so the structural transition can be seen as the splitting of the previously disordered ammonium into two independent positions, triggered by an reduction of the ND···Cl bond distances between the ND$_4$ molecules and the [FeCl$_5$·H$_2$O]$^{2-}$ units (see Figure 3). A detailed list with all the possible H-bonds can be consulted in Table 3.

The Fe···Fe distance along the zig-zag chain running along the *b*-axis is slightly longer than the observed at high temperature, with a value of 6.466(2) Å, while the interchain distance, which connects the Fe atoms in the *ac*-plane, is Fe···Fe 6.8061(13) Å, slightly longer compared with the high temperature one.

**Magnetic Structure**

The magnetic structure of (ND$_4$)$_2$[FeCl$_5$(D$_2$O)] was obtained from single-crystal neutron diffraction data taken at 2 K. is. A survey of reciprocal space was carried out with a series of *Q*-scans in the first Brillouin zone showing the appearance of superlattice reflections indexed by a propagation vector parallel to the *c*-axis, **k** = (0,0,$k_z$), with $k_z$ = 0.2288(4). This corresponds to a period of 30.2 Å in real space. In the following section, the magnetic structure is described by labeling the four Fe(III) atoms in the primitive unit-cell as Fe(1), Fe(2), Fe(3) and Fe(4) respectively with crystallographic coordinates (0.388, 0.249, 0.313), (0.119, 0.751, 0.813), (0.619, 0.751, 0.687) and (0.881, 0.249, 0.187). The magnetic moment [*m$_l$*(j)] for atoms at positions Fe(j) *(j=1 to 4)* in the unit-cell *l*, can be calculated by means of the Fourier expansion:

$$m_l(j) = \text{Re}(Sj)\cos\{2\pi \cdot [\mathbf{k} \cdot \mathbf{R}_l + \varphi(j)]\} + \text{Im}(Sj)\sin\{2\pi \cdot [\mathbf{k} \cdot \mathbf{R}_l + \varphi(j)]\}$$

where **R**$_l$, is the position vector of the unit-cell *l* with respect to the origin $\mathbf{R}_l = l_1\mathbf{a} + l_2\mathbf{b} + l_3\mathbf{c}$, where $l_i$ are integers, $\varphi(j)$ are relative phases (in fractions of 2π) and **Re**(S*j*) and **Im**(S*j*) are the real and imaginary parts of the Fourier vectors for each site *j*.

In order to determine the possible magnetic structures compatible with the crystallographic space group (*P*112$_1$/*a*) and with the **k** = (0,0,$k_z$) propagation vector, we have used representational analysis as described initially by Bertaut.[25] The decomposition of the magnetic representation (Γ) in the group of the wave-vector (little group) involves two one-dimensional irreducible representations (Γ$_1$ and Γ$_2$). In this symmetry group, only Fe(1)/Fe(2) on one hand, and Fe(3)/Fe(4) on the other, are related by symmetry elements, so that the orbit splits into two sets. For each of them, Γ=3Γ$_1$+3Γ$_2$, generating three sets of basis vectors for each representation (Table 4).

Magnetic structures compatible with a single representation (magnetic space group *P*112/*a*1'(0,0,*g*)00*s*) have been tested for all combinations of the real and imaginary parts of the Fourier vectors S(j) (i.e. describing all collinear and non-collinear magnetic arrangements). In all cases, the refinements lead to poor agreement factors (over 25 % $R_F^2$) and therefore these solutions can be excluded. In particular, the amplitude-modulated models systematically overestimate the Fe(III) magnetic moment, with amplitude maxima of 5.23(3) $\mu_B$, not physically meaningful. It deserves to be noted that none of these magnetic models would generate a macroscopic ferroelectric polarization, as observed experimentally.

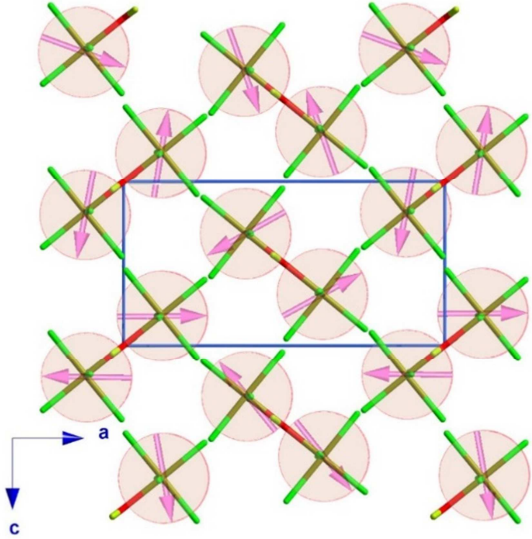

**Figure 4.** View along the *b*-axis of the superposition of the nuclear and magnetic structures for the combination of Γ1 and Γ2 irreducible representations. For the sake of clarity only [FeCl5(D2O)] units, have been represented. It should be noted that this combination of irreducible representations produces the same chirality for every spin chain running along the *c*-axis.

In order to fit successfully the experimental data, a combination of magnetic modes belonging to the $\Gamma_1$ and $\Gamma_2$ irreducible representations is required. The best solution was obtained with a cycloid model with moments mainly contained in the *ac*-plane and propagating along the *c*-direction (see Figure 4). This corresponds to the admixture in phase quadrature of a mode along the *a*-axis belonging to $\Gamma_1$ and a mode along the *c*-axis belonging to $\Gamma_2$; it should be noted that the *v* and *q* component of the $\Gamma_1$ and $\Gamma_2$ irreducible representations permit a small component along the *b*-axis. This solution has been confirmed by global optimization using a simulated annealing procedure.[21]

In the final refinements, the real and imaginary parts of the Fourier components S(j) were constrained to be equal, i.e. $m_{Fe}=|Re(Sj)|=|Im(Sj)|$, which corresponds to a spin cycloid with a circular envelope. The magnetic moments rotate within *a*-that includes the propagation vector and which forms an angle of 4.4(1)° with the *ac*-plane. A refinement using an elliptical envelope instead (Re(Sj)≠Im(Sj)), does not improve significantly the agreement factors: the $R_F^2$ reaches a value of 12.6 % and the difference between real and imaginary parts of the Fourier components is ca 0.3 $\mu_B$. The best fit between the observed and calculated intensities was obtained with phase angles of zero for Fe(1) and Fe(3) and $\mathbf{k}_z$/2 for Fe(2) and Fe(4); therefore Fe(1) and Fe(3) as well as Fe(2) and Fe(4) are strictly antiparallel, while the angle between the magnetic moments of Fe(1)-Fe(4) and Fe(2)-Fe(3) is ca 41.5° (see Figure 4). A full list of structural and magnetic parameters is given in Table 5, as well as the result of the magnetic refinement at 2K (see also Figure S3). The refined value of the Fe(III) magnetic moment, 3.805(2) $\mu_B$, is in agreement with those previously reported for the family of general formula $A_2[FeX_5(H_2O)]$, where A stands for an alkali metal or ammonium ion and X for a halide ion.[12] This value, slightly below the expected for a Fe(III) ion, indicates a non-negligible spin delocalization from the Fe(III) ions to the coordinated chloride atoms.[26]

**Table 4.** Magnetic moments of the content of a primitive cell deduced for the two possible irreducible representations ($\Gamma_1$ and $\Gamma_2$) and for the combination of both, for the magnetic sites: Fe(1)$_{4e}$ = (0.388, 0.249, 0.313), Fe(2)$_{4e}$ = (0.119, 0.751, 0.813), Fe(3)$_{4e}$ = (0.619, 0.751, 0.687) and Fe(4)$_{4e}$ = (0.881, 0.249, 0.187).

|  | Γ1 | Γ2 |
|---|---|---|
| $\mathbf{m}_{Fe(1)}$ | (u,v,w) | (u,v,w) |
| $\mathbf{m}_{Fe(2)}$ | ε(-u,-v,w) | ε(u,v,-w) |
| $\mathbf{m}_{Fe(3)}$ | (p,q,r) | (p,q,r) |
| $\mathbf{m}_{Fe(4)}$ | −ε(-p,-q,r) | −ε(p, q,-r) |

The symmetry operators relating the positions are: (1) = *x, y, z*; (2) = -*x*+1/2,-*y*+1,*z*-1/2; $\varepsilon$ = exp{-i·2π· 0.1136}, the common phase can be set to zero on the single representation, while for combined representations the phase difference between the basis vectors of the respective representations is π/2.

The admixture of two magnetic modes belonging to different irreducible representations lowers the point symmetry to *m* (*ab*-mirror plane) in perfect agreement with the emergence of an electrical polarization (**P**) in the *ab*-plane.

The presence of cycloids rotating in the *ac*-plane (with a small tilt of 4.4° along *b*) suggests that **P** should be preferentially directed along the *a*-axis, through the spin current mechanism induced via the inverse Dzyaloshinskii-Moriya interaction. The small tilt of the cycloid rotation plane towards *b*, is consistent with the weak electric polarization component along the *b*-axis previously observed.[11]

**Table 5.** Magnetic parameters of (NH4)2[FeCl5(H2O)] obtained from the refinement of the magnetic reflections recorded at 2 K.

| *a*(Å) | 13.5019(7) | *b*(Å) | 9.9578(5) | *c* (Å) | 6.9049(4) |
|---|---|---|---|---|---|
| α (°) | 90.00 | β (°) | 90.00 | γ(°) | 90.109(4) |
| k | (0,0,$k_z$) | $k_z$ | 0.2288(4) | $M_{Fe}$ ($\mu_B$) | 3.805(2) |
| φ(1)(2π) | 0.0 | φ(2)(2π) | 0.1136 | Ext. Model[a] | Anisotropic |
| $\phi_{Re}$(°) | 177.11(6) | $\phi_{Im}$(°) | -92.89(6) | $R_{exp}$(%) | 4.25 |
| $\theta_{Re}$(°) | 90.0 | $\theta_{Im}$(°) | 175.6(1) | $R_F$(%) | 12.8 |

$\phi$ and $\theta$ are the spherical angles of the Fourier components of the magnetic moment (real and imaginary).
[a] Fixed values from the structural refinement at 2K.

The symmetry of the magnetic structure is consistent with either a first-order transition or two successive transitions, the latter being consistent with the presence of two anomalies observed in the low temperature region of the heat capacity (ca 6.9 and 7.2 K). This scenario is also in agreement with experiments conducted by Mösssbauer spectroscopy [27] and muon spin relaxation,[28] both pointing to the occurrence of two distinct magnetic structures. This sequence of transitions suggests that two order parameters directed respectively along the *a*- or *c*- crystallographic axis (and belonging to $\Gamma_1$ and $\Gamma_2$), condense in turn at $T_{FE}$ and $T_N$. This corresponds to a transition on warming from cycloidal (T<$T_{FE}$) to collinear structure ($T_{FE}$<T<$T_N$), the latter involving a single irreducible representation and being compatible with the absence of electric polarization since this phase preserves inversion symmetry.[11] In fact, this would be reminiscent of the scenario observed in other compounds like $TbMnO_3$ or $MnWO_4$, where indeed such a sequence of two magnetic transitions is observed.[3, 4]

This hypothesis is reinforced by the temperature dependence of selected magnetic Bragg intensities such as (0,-3, $k_z$) and (0,0,-1-$k_z$) recorded in the temperature range of 2 - 12K (see Figure 5). The first one shows a continuous decrease with increasing temperature up to its complete disappearance at $T_N$. In contrast, the second presents a discontinuous change in the slope at ca. 6.9 K, in correspondence with the lowest temperature peak observed in heat capacity and with the onset of electric polarization.[11] However, the variation in intensity of these magnetic reflections does not allow to determine unambiguously the exact nature of the magnetic phase at $T_{FE}$<T<$T_N$. Moreover, the decrease of magnetic intensities in the vicinity of the paramagnetic region precludes the measurement of a full set of magnetic reflections.

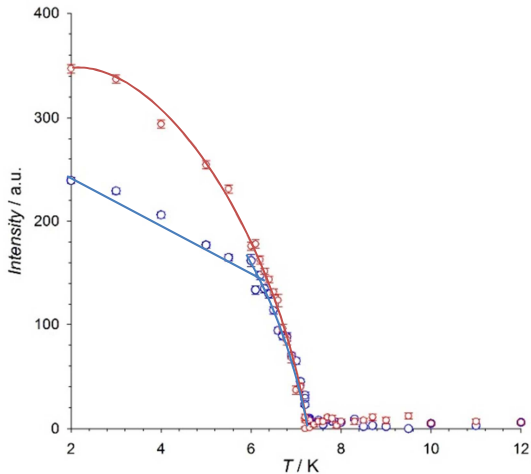

**Figure 5.** Temperature evolution of the (0,0,-1-$k_z$) and (0,-3, $k_z$) magnetic reflections, represented as blue and red circles, respectively. The change of curvature in the (0,0,-1.23) reflection could be associated to a magnetic phase transition from cycloid to collinear magnetic structure. The solid lines are guides to the eye.

The proposed magnetic structure at zero magnetic field can be explained by simple considerations regarding the exchange integrals, discussed using the notation already employed in ref 16. All magnetic exchange couplings are mediated by super-super exchange interactions, either through Cl-Cl or Cl-water pathways, and bound to be antiferromagnetic in nature. From previous DFT calculation in the $K_2[FeCl_5(H_2O)]$ compound,[29] $J_1$ that couples Fe(1)-Fe(3) and Fe(2)-Fe(4) ions, must be the strongest interaction due to the short halogen-oxygen together with the enhancement of the magnetic interaction due to the well oriented hydrogen bond. Indeed our magnetic configuration shows that the spins of these pairs of ions are strictly antiparallel. For the sake of simplicity, we assume $J_1$=1 for the rest of the discussion.

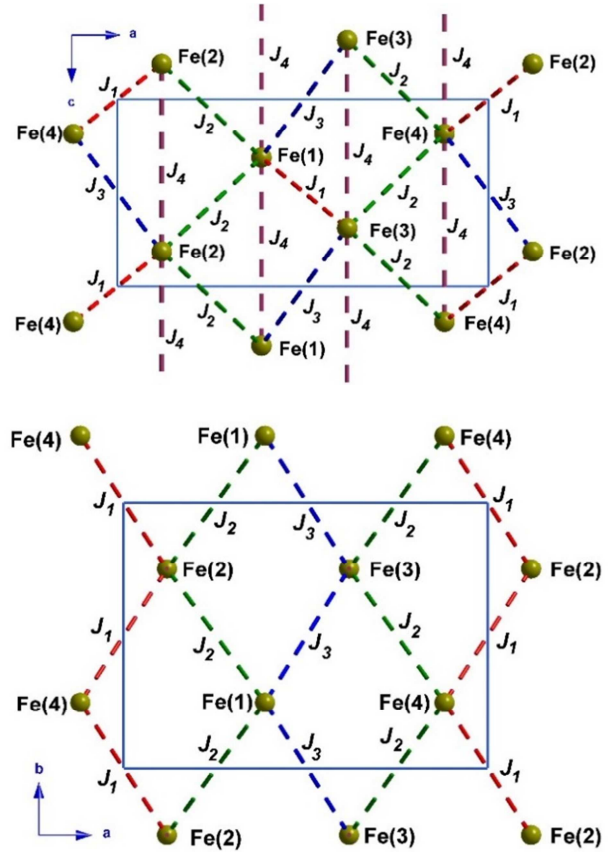

**Figure 6.** View along *b*- and *c*-axis of the four magnetic interactions using to model the magnetic ground state at 2K (see main text). The different exchange couplings have been represented in color dashed lines. For the sake of clarity only the Fe(III) ions have represented.

The next most relevant topological feature is the presence of buckled anisotropic triangular layers lying in the *bc*-plane, involving the $J_2$ and $J_4$ interactions (see Figure 6). This minimal set of parameters ($J_1$, $J_2$, $J_4$) allows to obtain a magnetic ground state which reproduces the experimental wave-vector ($k_z$=0.23), since for classical Heisenberg spins, the analytic solution of the anisotropic triangular lattice ($J_2/J_4$) is a spiral structure with $k_z = 2 \cdot acos[J_2/(2 \cdot J_4)]$. In the present case, one can derive $J_4 \approx 2/3 \cdot J_2$. It shall be noted that exact diagonalization of the first-ordered ground state shows that the minimum energy for any set $\{J_1, J_2, J_3, J_4\}$ is either at the Gamma point or along the $\Lambda$ symmetry line $(0,0,k_z)$ in CDML notation.[30]

The magnetic structure determined experimentally shows that the relative angles between spins on neighboring sites are slightly different from those obtained using $\{J_1, J_2, J_4\}$ alone. A more realistic solution matching closely the observed angles requires the addition of a weak $J_3$ coupling. By including $J_3$ however, the $J_4/J_2$ ratio must be renormalized in order to preserve the experimental wave-vector. For example, for $J_3 = 0.1$, and $J_4/J_2 \approx 0.75$, the magnetic structure can be fully reproduced. A weaker $J_3/J_4$ ratio compared to that reported for the $K_2[FeCl_5(H_2O)]$ compound seems compatible with the shorter $Cl(5)\cdots O(1w)$ and $Cl(3)\cdots Cl(4)$ distances mediating the $J_4$ coupling. The set of exchange couplings proposed here may not be a unique solution since some values were arbitrarily fixed, but their relative magnitude ($J_1 > J_2 > J_4 > J_3$) are consistent with the observed structure. Ab-initio calculations and/or measurements of magnetic excitations are required to provide quantitative values of these exchange couplings.

## Conclusion

With the ultimate goal of establishing the mechanism at the origin of multiferroicity in the title compound, we have carried out crystal and magnetic structure studies as a function of the temperature.

In previous works, the transition identified at 79 K by heat capacity measurements was hypothetically attributed to a structural phase transition, but all efforts to determine the crystal structure below this temperature led to solutions indistinguishable from the high temperature structure. In this work, we have taken advantage of the contrast provided by neutron diffraction for the precise determination of the positions of light atoms (in this case mainly the hydrogen/deuterium atoms of the ammonium molecules) to characterize this structural phase transition, which is ultimately explained by a blocking of the ammonium counterions.

The space group analysis at low temperature has shown that the title compound crystalizes in the $P112_1/a$ space group. The centrosymmetric character of the space group precludes ferroelectric polarization, as in fact is observed in the previously reported pyroelectric current measurements just below the structural phase transition.[11]

The compound becomes magnetically ordered at 7.2 K ($T_N$) and becomes ferroelectric below 6.9 K ($T_{FE}$), with a spontaneous electric polarization presenting two different components: a strong one along the $a$-axis and a weak one along the $b$-axis, with a value ten times lower. Our magnetic neutron diffraction unambiguously shows that at 2K, an incommensurate cycloidal spin structure is stabilized at zero magnetic field. The magnetic moments are contained mostly in the $ac$-plane (ca 4° tilt along $b$) and propagate along the $c$-direction. The symmetry lowering at 2K confirms that the ferroelectricity is a direct consequence of the complex magnetic ordering and suggests that the *inverse* Dzyaloshinskii-Moriya mechanism is at play in the magnetoelectric coupling. This is in sharp contrast with all other compounds of the series ($A_2[FeX_5(H_2O)]$, where A stands for an alkali metal and X is an halogen atom) presenting commensurate collinear antiferromagnetic structures and not ferroelectric. This highlights that subtle chemical modifications produce remarkable changes in the physical behavior of this class of compounds. Specifically, the substitution of the alkali-metal for an ammonium molecule modifies the exchange coupling interactions giving rise to a frustrated magnetic structure. We have provided a minimalistic model based on a set of four isotropic exchange interactions which explains well the observed magnetic state. Strong competition within anisotropic triangular layers seems to be responsible for the observed behavior. It is likely that external parameters such as magnetic field or pressure influence drastically the spin state and potentially lift the frustration, as recently suggested from magnetoelectric measurements under external magnetic field. Further neutron diffraction experiments under external magnetic field are needed in order to elucidate the different magnetic models and determine the mechanism of magnetoelectric coupling in the different regions of the phase diagram.

## Notes and references

**Acknowledgements**
Partial funding for this work is provided by the Ministerio Español de Ciencia e Innovación through projects MAT2010-16981, MAT2011-27233-C02-02. JARV acknowledges CSIC for a JAEdoc contract.
We are grateful to the Institut Laue Langevin for the neutron beam-time allocated. The authors are especially grateful to Dr. C. Ritter (ILL) for the access to the D2B diffractometer for the high resolution low temperature neutron measurement.

**Corresponding Authors**
* OF fabelo@ill.fr  and LC chapon@ill.fr

**Supporting Information**
Single crystal X-ray crystallographic data at RT and 50 K, detail of the heat capacity curves of $(ND_4)_2[FeCl_5(D_2O)]$ compound, high-resolution neutron powder diffraction pattern at 45 K and Rietveld refinement using two different space groups, squared magnetic structure factors observed *versus* calculated for D9 instrument at 2K.

**Abbreviations**
**1_HT**, Compound **1** High Temperature; **1_LT**, Compound **1** Low Temperature; KPI is the acronym of Kitaigorodskii Packing Index; CDML is the acronym of Cracknell, A. P., Davies, B.L., Miller, S. C. and Love, W. F.

# Supporting information for:

# Magnetically-induced ferroelectricity in the (ND$_4$)$_2$[FeCl$_5$·D$_2$O] molecular compound.

José Alberto Rodríguez-Velamazán,[1,2] Óscar Fabelo,[2*] Ángel Millán,[1] Javier Campo,[1]

Roger Johnson,[3] Laurent Chapon.[2*]

[1] Instituto de Ciencia de Materiales de Aragón (ICMA), CSIC – Universidad de Zaragoza, 50009 Zaragoza, Spain.

[2] Institut Laue-Langevin, 38042 Grenoble Cedex 9, France.

[3] Univ Oxford, Dept Phys, Clarendon Lab, Oxford OX1 3PU, England

**Table S1.** Experimental parameters and main structural crystallographic data for the studied compounds determined from single crystal X-ray diffraction.

| Formula | [NH$_4$]$_2$[Fe$^{III}$Cl$_5$(H$_2$O)] | |
|---|---|---|
| Empirical Formula | Cl$_5$H$_{10}$FeN$_2$ | |
| Mr (g·mol$^{-1}$) | 287.20 | |
| Temperature (K) | 293(2) | 50(2) |
| $\lambda$(Å) | 0.71073 | 0.71073 |
| Crystal system | Orthorombic | Orthorombic |
| Space group (No.) | *Pnma* (62) | *Pnma* (62) |
| Crystal size (mm) | 0.08 × 0.08 × 0.06 | 0.08 × 0.08 × 0.06 |
| $a$ (Å) | 13.728(5) | 13.5088(6) |
| $b$ (Å) | 9.934(5) | 9.9408(4) |
| $c$ (Å) | 7.040(5) | 6.9070(4) |
| $\alpha$(º) | 90.00 | 90.00 |
| $\beta$(º) | 90.00 | 90.00 |
| $\gamma$(º) | 90.00 | 90.00 |
| $V$ (Å$^3$) | 960.0(9) | 927.53(8) |
| $Z$ | 4 | 4 |
| $\rho_c$ (g·cm$^{-3}$) | 1.987 | 2.057 |
| Meas. Reflections/ (Rint) | 26820/(0.0442) | 16496/(0.0931) |
| Indep. ref. [I > 2σ(I)] | 943 | 923 |
| Parameters/ restraints. | 69/ 4* | 69 / 4* |
| Hydrogen treatment | Refall | Refall |
| Goodness of fit | 1.139 | 1.132 |
| Final R indices [I > 2σ(I)]: $R_1$ / $wR_2$ | 0.0387/ 0.0593 | 0.0554/0.0803 |
| R indices (all data): $R_1$ / $wR_2$ | 0.0768/0.0694 | 0.1011/0.0906 |

*The hydrogen atoms where refined using a DFIX command with a N-H bond distance of 0.90(3)

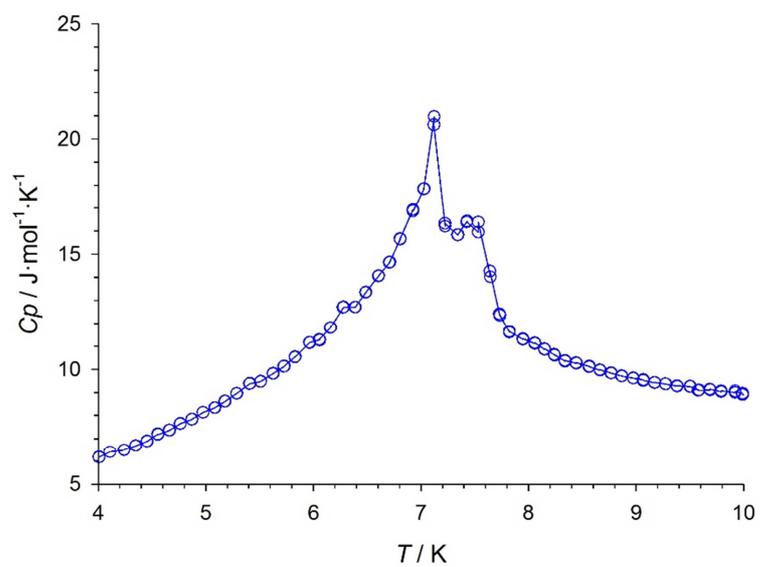

**Figure S1.** Details of the heat capacity curves of $(ND_4)_2[FeCl_5(D_2O)]$ compound in the low temperature range. The two different phase transitions observed (long range magnetic ordering at ca 7.4 K and ferroelectric ordering at ca 7.1K ) are in agreement with those previously characterized for the non-deuterated compound.

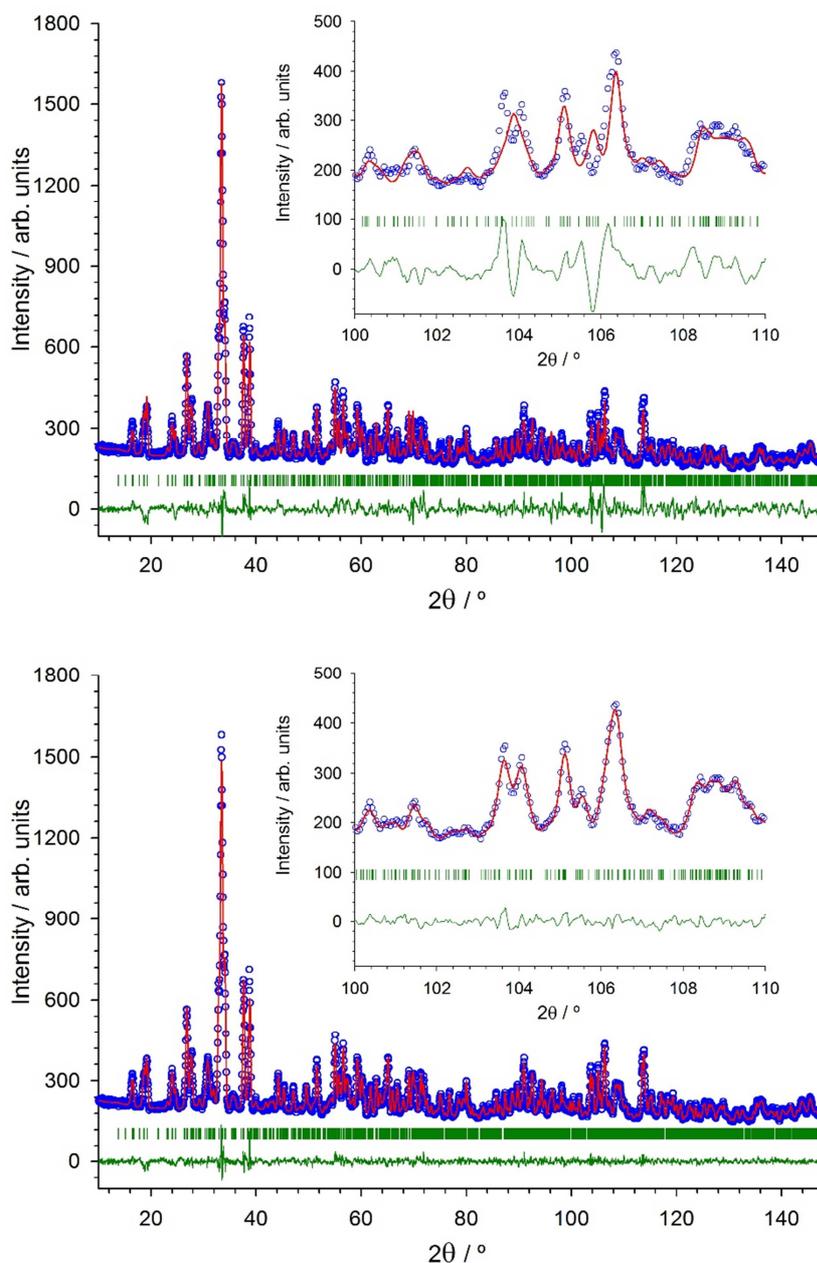

**Figure S2.** Neutron diffraction pattern of $(ND_4)_2[FeCl_5(D_2O)]$ compound collected at 45K using D2B high resolution instrument with $\lambda$= 1.5942 Å. The Rietveld refinements were done in the space group *Pnma* (up) and in *P*112$_1$/*a* (bottom), the insets show the splitting due to the nuclear phase transition from orthorhombic to monoclinic system. The experimental data have represented as blue circles, the calculated curve as solid red line and the difference between them as solid green line. The green vertical lines represent the Bragg positions for the selected space group.

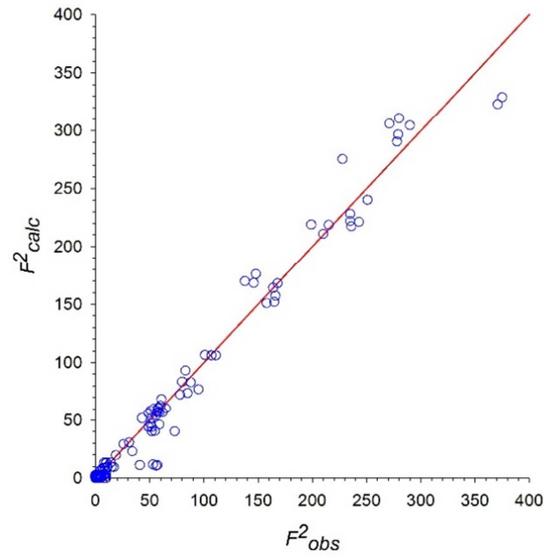

**Figure S3.** Squared magnetic structure factors observed *versus* calculated for D9 instrument at 2K. The refinement has been done with 127 magnetic reflections, with an agreement factor of $R_F$ = 12.8%.